\documentclass[usenatbib]{mn2e}
\usepackage{natbib}
\usepackage{amsmath,amssymb}
\usepackage{epsfig}
\usepackage{color}
\usepackage{graphicx}

\input{macros.tex}
\def\rhodm{\rho_{\rm DM}}

\title[Our Galaxy's dark halo]
{The distribution function of the Galaxy's dark halo}

\author[Binney \& Piffl]{J.~Binney\thanks{E-mail: binney@physics.ox.ac.uk} and T.~Piffl
\\
Rudolf Peierls Centre for Theoretical Physics, Keble Road, Oxford OX1 3NP, UK}

\begin{document}
\date{Draft, \today}
\pagerange{\pageref{firstpage}--\pageref{lastpage}} \pubyear{2015}
\maketitle
\label{firstpage}
\begin{abstract}
Starting from the hypothesis that the Galaxy's dark halo responded
adiabatically to the infall of baryons, we have constructed a self-consistent
dynamical model of the Galaxy that satisfies a large number of observations,
including measurements of gas terminal velocities and masers, the kinematics
of a 180\,000 giant stars from the RAVE survey, and star count data from the
SDSS. The stellar disc and the dark halo are both specified by distribution
functions (\df s) of the action integrals. The model is obtained by extending
the work of \citeauthor*{PifflPenoyreBinney2015} from the construction of a
single model to a systematic search of model space.  Whereas the model of
\citeauthor*{PifflPenoyreBinney2015} violated constraints on the
terminal-velocity curve, our model respects these constraints by adopting a
long scale length $R_\d=3.66\kpc$ for the thin and thick discs. The model is,
however, inconsistent with the measured optical depth for microlensing of
bulge stars because it attributes too large a fraction of the density at
$R\la3\kpc$ to dark matter rather than stars. Moreover, it now seems likely
that the thick disc's scale-length is significantly shorter than the model
implies.  Shortening this scale-length would cause the constraints from the
rotation curve to be violated anew.  We conclude that we can now rule out
adiabatic compression of our Galaxy's dark halo.

\end{abstract}

\section{Introduction}
Over the coming decade models of our Galaxy will play a crucial role in the
extraction of physical understanding from the data that are currently being
collected by a series of large surveys \citep[e.g.][]{Binneyreview}. The models
of choice are fully dynamical in the sense that they take full account of the
equations of motion of the Galaxy's constituent particles and of the laws of
gravitation.

An important class of such models represent the Galaxy by a large number of
particles. The model is then specified by the phase-space
coordinates, masses and other properties of the particles. Specifications of
this kind are highly redundant in the sense that on integrating the equations
of motion for a dynamically insignificant time, such as $10\Myr$, all the
phase-space coordinates change while the model remains the same.  This
redundancy is associated with great complexity in the sense that millions of
particles are required to achieve even mediocre spatial resolution in the
model, so (a) at least 10 million real numbers are required to specify a
model, and (b) given these numbers it is a totally non-trivial exercise to
characterise the model in a meaningful way. Moreover, as a consequence of
this complexity, the models are typically computationally expensive to
produce and it is unlikely to prove feasible to fit such a model to the wide
range of very detailed data for our Galaxy that are becoming available.

Models of a very different class are specified by a distribution function
(\df) $f_\alpha(\vJ)$ for each population $\alpha$ of constituent particles.
Each \df\ is an analytic function of three constants of orbital motion, which
it is convenient to choose to be the actions $J_i$, and specifies the
probability density of particles of type $\alpha$ in the three-dimensional
region of phase space associated with a given set $\vJ$ of action values. The
legitimacy of assuming that \df s depend only on $\vJ$ is assured by Jeans'
theorem \citep{GalacticDynamics}. The model is specified by giving the values of the
parameters $\vp$ that appear  $f_\alpha$ alongside the actions.
Hence the model is specified by the values of $\lta100$ numbers, rather
than millions, and it is computationally feasible to adjust the parameters to
optimise the agreement between the model and the data. Moreover, the
parameters can be chosen to be approximately equal to intuitive properties of
the population, such as its radial scale length, radial and vertical velocity
dispersions in the solar neighbourhood, etc.

In a series of papers \cite{Binney2010,Binney2012b,Binney2014b,Piffl2014b} we
have demonstrated the effectiveness of a particular form of \df\ for disc
stars.  \cite{Binney2012b} fitted this \df\ to the kinematics of stars in the
Geneva-Copenhagen Survey \citep[][hereafter
GCS]{GCS2004,Holmberg2007,Casagrande2011}, which lie within $\sim120\pc$ of
the Sun. The resulting model was then used to predict the kinematics of stars
in the RAdial Velocity Experiment \citep[][hereafter
RAVE]{RAVE_DR1,RAVE_DR4}, which extend to $\sim2.5\kpc$ of the Sun
\citep{Binney2014b}. The ability of the model to reproduce data very
different from that used in the model's specification is remarkable in view
of the limited extent to which the thick disc contributes to the GCS data.
\citet[][hereafter P14]{Piffl2014b} showed that minor adjustments to the
parameter values in \cite{Binney2012b} enabled the model to fit
simultaneously the RAVE data and data from the Sloan Digital Sky Survey in
\citet[][hereafter \citetJuric]{Juric2008}, and yielded the currently strongest constraints on the mass
of dark matter within the solar radius, $R_0$.

Fundamentally stars form a multi-dimensional continuum spanned by mass, age,
and chemistry. Models need to take cognisance of the masses, ages and
chemical compositions of stars because these intrinsic variables determine a
star's absolute magnitudes in the various observing bands, and thus the
probability that a star will be included in any given data set.  The work
summarised above handled these complexities very crudely by supposing that
our Galaxy comprises a continuum of stellar populations that differ only in
age. Even within this rough approximation the analysis was unnecessarily
crude. \cite{SandersBinney2015a} moved the use of analytic \df s to a new
level of sophistication by (a) introducing metallicity [Fe/H] as an argument
of the \df\ of disc stars, and (b) using isochrones to compute the number of
stars the model predicts will be observed in given surveys.
\cite{SandersBinney2015a} fitted their ``extended \df'' for the disc to the
GCS and compared the resulting predictions for the more distant G dwarfs in
the Sloan Extension for Galactic Understanding and Exploration
\citep[][hereafter SEGUE]{SEGUEpaper}.

Although much remains to be done, as a result of the work just summarised, we
now have a reasonable understanding of the phase-space distribution of disc
stars -- the extensions that seem most urgent are to compare the model's
predictions with data for stars that never come close to the Sun, and to
model the way non-stationary features such as the Galactic bar and spiral
structure modify the model's static \df.

Our Galaxy is by no means comprised of just stars: overall more than 90
percent of its mass is thought to be contributed by dark-matter particles,
and from the RAVE data using an analytic \df\ for the disc, P14 concluded
that even interior to the Sun, where baryons make their largest contribution
to the overall mass, more than half the matter is dark. 

In the work with analytic \df s so far mentioned, dark matter was included
merely by adding to the gravitational field generated by stars and gas a
hypothetical contribution from the dark halo. The density distribution of the
dark halo that was used by P14 and similar studies was inspired by
simulations of cosmological clustering that did not include baryons, and
hence side-stepped the complex physics that baryons bring into play. In these
simulations the density profiles of dark halos are well approximated by a
simple functional form, the so-called NFW profile, regardless of the mass or
assembly histories of individual halos \citep{NFW1997}. Given that we have
yet to actually detect any dark-matter particles, and dark matter strongly
dominates the Galaxy's overall mass budget, real dark halos are widely
modelled as spherical or mildly spheroidal version of the profile.

The intellectual basis for this procedure is weak, however, because even if a
given dark halo conformed to the NFW profile before the bulk of the baryons
flowed in and formed stars, it would be a remarkable coincidence if it
conformed to this profile now. Indeed, if the baryons accumulated over many
dynamical times rather than in a few massive infall events then the actions
of the dark-matter particles will be invariant as the Galaxy grows, and it
will be the \df\ of the dark matter, not its spatial density profile, which
will be invariant.  The existence of an old thin disc favours the hypothesis
that baryons accumulate quiescently.

This line of argument motivated \cite[][hereafter
PPB]{PifflPenoyreBinney2015} to open a new chapter in the use of analytic \df
s by specifying the dark halo through its \df\ rather than its spatial
density distribution. Hence the Galaxy model they fitted to RAVE data was
based on analytic \df s for both the stars and the dark matter, and the
spatial distributions of these components were determined by iteratively
solving for the gravitational potential that these components jointly
generate in concert with smaller, specified contributions from gas and an
axisymmetrised version of the Galactic bulge/bar.  PPB took as their point of
departure the model P14 had fitted to RAVE and \citetJuric\ data. This model
provided the complete \df\ of the disc and a particular NFW profile. PPB
adopted the disc \df, and for the \df\ of the dark halo used one that
self-consistently generates the given NFW profile {\it in the absence of the
disc}. PPB found that when this dark halo cohabits with the disc, the disc's
gravitational field contracts the central portion of the dark halo
sufficiently to cause the final model to be inconsistent with the measured
circular speed $\vc$ at $R\lta6\kpc$.

This conflict between data and the first model of our Galaxy to specify the
dark halo through its \df, can be addressed in two extreme ways: one can
modify the \df\ of either the disc or the dark halo, leaving the other alone.
Modification of the halo's \df\ would imply that the infall of baryons was
sufficiently unsteady to violate adiabatic invariance of the actions of halo
particles. Here we ask whether this conclusion can be avoided by leaving the
functional form of the dark halo's \df\ alone and seeking a disc \df\ that is
consistent with the observational data when the disc cohabits with a dark
halo of this form, and the previously assumed quantities of gas and bulge
stars. We will show that with current data it is possible to model the RAVE
and SEGUE data successfully within constraints set by measurements of $\vc$
by increasing the scale length of the disc from $\sim2.5\kpc$ to
$\sim3.5\kpc$, but find that the resulting model then predicts too little
microlensing of bulge stars.

A key achievement of this paper is the development of a practical technique
for searching a multi-dimensional model space for a model that satisfies
observational constraints on the space density and kinematics of stars in the
extended solar neighbourhood, and constraints from gas, masers and Sgr A* on the
Galaxy's rotation curve. In forthcoming work we plan to use this technique to
obtain fully dynamical models of our Galaxy that are consistent with all
available observations.

In Section~\ref{sec:constraints} we list our observational inputs. In
Section~\ref{sec:DFs} we specify the functional forms that define our models.
In Section~\ref{sec:algorithm} we explain how we fit the data.
Section~\ref{sec:results} describes the best-fit model.
Section~\ref{sec:discuss} discusses the implications of the model and in
Section~\ref{sec:conclude} we sum up and draw conclusions.

\section{Observational inputs}\label{sec:constraints}
We use the same observations as in P14, apart from dropping the
constraints for the dark halo. We assume a distance of the Sun to the
Galactic centre (GC), $R_0$, to be 8.3 kpc \citep[e.g.][]{Gillessen2009b,
McMillan2011, Schoenrich2012}, the distance of the Sun above the Galactic
plane, $z_0$, to be 14 pc \citep{BinneyGS1997} and the solar motion with
respect to the local standard-of-rest (LSR), $\vv_\odot$, to be
$(11.1,12.24,7.25)\kms$ \citep{Schoenrich2010}.

Our most important inputs are (i) the vertical  profile of stellar density
above the Sun determined by \citetJuric, and
(ii) the kinematics of $\sim180\,000$ giant stars in RAVE.

\subsection{Vertical density profile from SDSS}

We assume that the population from which the RAVE sample is drawn is
identical to that studied by \citetJuric, who measured its vertical density
profile by means of a main-sequence colour-magnitude relation. We use the
data points shown in the middle panel of their Figure~15, which shows results
from M dwarf stars in the colour range $0.70 < r-i < 0.80$. Similar to RAVE,
this sample should carry only weak biases in metallicity and age. Rather
than correcting the data for the effects of Malmquist bias and binarity as
\citetJuric\ did, P14 imposed these effects on the model. Since this step
had a negligible effect on their results, we omitted it.

We decomposed the \citetJuric\ density profile into contributions from the
disc and stellar halo. This decomposition implied that at
$(R,z)=(R_0,0.5\kpc)$, the density of the stellar halo is $0.0056$ times the
density of the disc. The \df\ of the stellar halo is subsequently always
normalised to produce this ratio of densities at $(R_0,0.5\kpc)$.

\subsection{Kinematics from RAVE}

 For the kinematics we use the stellar parameters and distance estimates in
the fourth RAVE data release \citep{RAVE_DR4}. 
We define eight spatial bins in the $(R,z)$ plane. Four bins for stars inside
the solar cylinder with $R_0-1\kpc < R < R_0$ and $|z|$ in
[0,0.3],[0.3,0.6],[0.6,1.0] or [1,1.5] kpc. The other four bins cover the
same $z$ ranges but cover the regions 1 kpc outside the solar cylinder, i.e.\
$R_0 < R < R_0 + 1\kpc$. After sorting the stars into these bins, we compute
the velocity distributions predicted by the \df\ at the mean $(R,z)$
positions (barycentre) of the stars in each bin. For each bin we have a
histogram for each component of velocity, so we accumulate $\chi^2$ from 24
histograms.  Throughout this work we compute velocities in the coordinate
system that \citet{Binney2014b} found to be closely aligned with the velocity
ellipsoid throughout the extended solar neighbourhood -- this system is quite
closely aligned with spherical coordinates. We denote the velocity component
along the long axis of the velocity ellipsoid -- pointing more or less
towards the Galactic centre -- with $V_1$, the azimuthal component with
$V_\phi$, and the remaining component with $V_3$.

The resulting model distributions cannot be directly compared to the observed
distributions, because the latter are widened by errors in the velocity and
parallax estimates. We fold the model distributions with the average velocity
uncertainties of the bin's stars to obtain $N_{\rm bary}(V_i)$. The
distortions arising from the parallax error are less straight forward to
introduce: following \citet{Binney2014b} we create a Monte Carlo realisation
of a given \df\ by randomly assigning to each star in our RAVE sample a new
``true'' distance according to its (sometimes multi-modal) distance pdf, and
a new ``true'' velocity according to the model velocity distribution at this
position. With these new phase space coordinates we compute new observed
line-of-sight velocities and proper motions. These are finally equipped with
random observational errors. Using the original catalogue distances, we then
compute new realistically distorted velocity distributions, $N_{\rm
MC}(V_i)$, based on the \df\ that can be compared directly to the original
RAVE distributions in a number of spatial bins. We minimise the Poisson noise
in $N_{\rm MC}(V_i)$ by choosing 100 new velocities for each star. This
procedure is computationally expensive and the distortions vary only weakly
for reasonable choices of the \df\ parameters. To speed up the process, we
store the ratio $N_{\rm bary}(V_i)/N_{\rm MC}(V_i)$ for a \df\ that is
already a good match of the RAVE data. Examples of these ratios, which are
near unity in the core of the distribution but fall to $<0.2$ in the wings,
are shown in the lower panels of Fig.~4 of P14. These
ratios are then used to correct all \df\ predictions before they are compared
with the data.

Our model
selection involves computing the corresponding velocity histograms at the
barycentre of each bin, and optimising the fit between the data and these
histograms after the latter have been modified to allow for the impact of errors in
the measurements of velocity and distance.

\subsection{Gas terminal velocities}
The distribution of \HI\ and CO emission in the longitude-velocity plane yield
a characteristic maximum (``terminal'') velocity for each line of sight
\citep[e.g.][\S9.1.1]{GalacticAstronomy}. The terminal velocities are related
to the circular speed $\vc(R)$ by
\begin{equation}
 \begin{split}
  v_\mathrm{term}(l) &= v_\mathrm{c}(R) - v_\mathrm{c}(R_0)\sin l \\
                     &= v_\mathrm{c}(R_0\sin l) - v_\mathrm{c}(R_0)\sin l.
 \end{split}
\end{equation}
We use the terminal velocities $v_{\rm term}(l)$ from \citet{Malhotra1995}.
Following \citet{Dehnen1998} and \citet{McMillan2011} we neglect data at
$\sin l < 0.5$ in order not to be influenced by the Galactic bar, and we
assume that the ISM has a Gaussian velocity distribution of dispersion
7~km\,s$^{-1}$.
\subsection{Maser observations}
\citet{Reid2014} presented a compilation of 103 maser observations that
provide precise 6D phase space information. Since masers are associated with
young stars their motions should be very close to circular around the GC. We
again assume an intrinsic velocity dispersion of $7\kms$ and no lag against
the circular speed \citep{vanderKruit1984,McMillan2010}. For the likelihood
computation we neglected 15 sources that were flagged as outliers by
\citet{Reid2014} and also all sources at $R < 4\kpc$. The latter is again to
prevent a bias by the Galactic bar. To assess the likelihood of a maser
observation, we predict the observed velocities (line-of-sight velocity,
proper motions) as functions of heliocentric distance and then integrate the
resulting probability density along the line-of-sight.
\subsection{Proper motion of SgrA*}
\citet{Reid2004} measured the proper motion of the radio source SgrA* in the
GC to be
\[
 \mu_\mathrm{SgrA^\star} = -6.379 \pm 0.024~\mathrm{mas\,yr}^{-1}.
\]
This source is thought to be associated with the super-massive black hole that
sits in the gravitational centre of the Milky Way with a peculiar velocity below
$1\kms$. Hence this measurement reflects the solar motion with respect to
the GC.
\section{Model definitions}\label{sec:DFs}
\subsection{The stellar disc}
The functional form of the stellar disc's \df, which is made up of
contributions from the thick disc and each coeval cohort of thin-disc stars,
is unchanged from PPB, even though \cite{SandersBinney2015a} have strictly
speaking rendered that form obsolete. The \df\ segments the disc by age: the
oldest stars are represented by a \df\ for the ``thick disc''. The thick
disc's \df\ is a ``quasi-isothermal'' component \citep{BinneyMcMillan2011}.
The \df\ of such a component is
\[\label{eq:qi}
 f(J_r,J_z,L_z)=f_{\sigma_r}(J_r,L_z)f_{\sigma_z}(J_z,L_z),
\] 
where $f_{\sigma_r}$ and $f_{\sigma_z}$ are defined to be
\[\label{planeDF}
 f_{\sigma_r}(J_r,L_z)\equiv \frac{\Omega\Sigma}{\pi\sigma_r^2\kappa}
 [1+\tanh(L_z/L_0)]\e^{-\kappa J_r/\sigma_r^2}
\]
and
\[\label{basicvert}
 f_{\sigma_z}(J_z,L_z)\equiv\frac{\nu}{2\pi\sigma_z^2}\,
 \e^{-\nu J_z/\sigma_z^2}.
\]
Here $\Omega(L_z)$, $\kappa(L_z)$ and $\nu(L_z)$ are, respectively, the
circular, radial and vertical epicycle frequencies of the circular orbit with
angular momentum $L_z$, while
\[\label{eq:defsSigma}
 \Sigma(L_z)=\Sigma_0\e^{-\Rc/\Rd},
\]
where $\Rc(L_z)$ is the radius of the circular orbit, determines the surface
density of the disc: to a moderate approximation the surface density at
Galactocentric distance $R$ can be
obtained by using for $L_z$ in equation (\ref{eq:defsSigma}) the angular
momentum $L_z(R)$ of the circular orbit with radius $R$. 

In equation (\ref{planeDF}) the factor containing tanh serves to eliminate
retrograde stars; the value of $L_0$ controls the radius within which
significant numbers of retrograde stars are found, and should be no larger
than the circular angular momentum at the half-light radius of the bulge/bar.
Provided this condition is satisfied, the results for the extended solar
neighbourhood presented here are essentially independent of $L_0$.

The \df\ of the thin disc is taken to be a superposition of quasi-isothermal
\df s, one for the stars of each age $\tau$. The velocity-dispersion
parameters $\sigma_r$ and $\sigma_z$ above are functions $\sigma_r(L_z,\tau)$
and $\sigma_z(L_z,\tau)$ of angular momentum and age. They control the radial
and vertical velocity dispersions of the stars of age $\tau$ and are
approximately equal to them at $\Rc$.  Given that the scale heights of
galactic discs do not vary strongly with radius \citep{vanderKruit1981},
these quantities must increase inwards, and we assume this dependence on
$\Rc$ is exponential.  We take the growth with age of the velocity
dispersions of a coeval cohort of thin-disc stars from the work of
\citet{Aumer2009}. With these assumptions the velocity-dispersion parameters
are given by
\begin{eqnarray}\label{sigofLtau}
 \sigma_r(L_z,\tau)&=&\sigma_{r0}\left(\frac{\tau+\tau_1}{\tau_{\rm m}+\tau_1}\right)^\beta\e^{(R_0-\Rc)/R_{\sigma,r}}\nonumber\\
 \sigma_z(L_z,\tau)&=&\sigma_{z0}\left(\frac{\tau+\tau_1}{\tau_{\rm m}+\tau_1}\right)^\beta\e^{(R_0-\Rc)/R_{\sigma,z}}.
\end{eqnarray}
Here $\sigma_{z0}$ is the approximate vertical velocity dispersion of local
stars at age $\tau_{\rm m}\simeq10\Gyr$, $\tau_1$ sets velocity dispersion at
birth, and $\beta\simeq0.33$ is an index that determines how the velocity
dispersions grow with age. The radial scale-lengths on which the velocity
dispersions decline are $R_{\sigma,i}$, and  a constant scale height would be
follow if $R_{\sigma,z} \sim 2\Rd$.

We assume that the star-formation rate in
the thin disc has decreased exponentially with time, with characteristic time
scale $t_0$, so the thin-disc \df\ is
\[\label{thinDF}
 f_{\rm thn}(J_r,J_z,L_z)=\frac{\int_0^{\tau_{\rm m}}\rd\tau\,\e^{\tau/t_0}
 f_{\sigma_r}(J_r,L_z)f_{\sigma_z}(J_z,L_z)}{t_0(\e^{\tau_{\rm m}/t_0}-1)},
\]
where $\sigma_r$ and $\sigma_z$ depend on $L_z$ and $\tau$ through equation
(\ref{sigofLtau}). We set the normalising constant $\Sigma_0$ that appears in
equation (\ref{eq:defsSigma}) to be the same for both discs and use for the
complete \df
\[
 f_{\rm disc}(J_r,J_z,L_z)=f_{\rm thn}(J_r,J_z,L_z) +
 F_{\rm thk}f_{\rm thk}(J_r,J_z,L_z),
\]
where $F_{\rm thk}$ is a parameter that controls the fraction $(1+F_{\rm
thk}^{-1})^{-1}$ of stars that belong to the thick disc. The values of the
parameters for our final model are given in Table~\ref{tab:discDF}.

We followed PPB in imposing a lower limit of $1\kpc$ on the value of
$\Rc(J_\phi)$ at which the epicycle frequencies $\kappa(J_\phi)$ and
$\nu(J_\phi)$ are evaluated for use in the \df.

\subsection{DF of the stellar halo}\label{sec:starhalo}

As in P14, we include the contribution of a stellar halo when fitting the
kinematics of RAVE stars, which include a small but non-negligible population
of  stars that are  identifiable as halo stars by their low or even negative
values of the azimuthal velocity $v_\phi$.  Including the \df\ of the stellar
halo prevents the fitting routine distorting the thick disc in an attempt to
account for the presence of halo stars in the sample.

The density of the stellar halo is generally thought to follow a power-law in
Galactocentric radius, i.e.\ $\rho_{\rm halo} \propto r^{-\alpha}$, with the
power-law index $\alpha \simeq 3.5$ \citep[e.g.][\S10.5.2]{GalacticAstronomy}.
We can model such a configuration using the following form of the
(un-normalised) \df\
\citep{Posti2014}
\[\label{eq:haloDF1}
f(\vJ)  = g(\vJ)^{3.5}\exp\{-[g(\vJ)/g_{\rm max}]^4\},
\]
where
\[
g(\vJ)\equiv J_r + \gamma_1|L_z| + \gamma_2 J_z + J^{\rm(s)}_{\rm core}
\]
with $\gamma_1 = 0.937$, $\gamma_2 = 0.682$, $J^{\rm(s)}_{\rm core} = 200\kms\kpc$ and
$g_{\rm max} = 2.5\times10^5\kms\kpc$. These choices of $\gamma_1$ and
$\gamma_2$ make the stellar halo approximately spherical.  The RAVE data
alone are not well suited to constraining the stellar halo, so we defer this
exercise to a later paper (Das \& Binney in preparation). We include the
stellar halo only to prevent distortion of the thick disc that is
fitted to the data.  Our complete total stellar \df\ is
\begin{equation}
 f(J_r,J_z,L_z) = f_{\rm disc}(J_r,J_z,L_z) + F_{\rm halo}f_{\rm halo}(J_r,J_z,L_z)
\end{equation}
with $F_{\rm halo}$ chosen so $\rho_{\rm halo}/\rho_{\rm disc}=0.0056$
$0.5\kpc$ above the Sun to be consistent with the \citetJuric\ data as
explained at the start of Section~\ref{sec:constraints}.

\subsection{The dark halo}

\cite{Posti2014} found a simple \df\ $f(\vJ)$ that
self-consistently generates a spherical model that has almost exactly an NFW
profile. This model is essentially isotropic near its centre, and becomes
mildly radially biased beyond its scale radius. PPB extended this \df\ so
they could explore how halos with different velocity anisotropies respond to
the infall of baryons. They reported results for three model dark halos, one
radially biased, one tangentially biased and the original nearly isotropic
model of Posti et al. Here we consider only the radially biased case, which
most closely resembles the halos that form in simulations that include only
dark matter. The PPB \df\ is defined  in terms of an approximately homogeneous
function of the actions
\begin{equation} \label{eq:hJ}
 h(\vJ)\equiv {1\over A}J_r + \frac{\Omega_\phi}{B\kappa} (|J_\phi| +
 J_z),
\end{equation}
 where $A$ and $B$ are given by equations (6) and (7) of PPB with $b=8$ to ensure
 radial anisotropy:
\[\begin{split}
A&=4.5+3.5\tanh^2\left({|L_z|+J_z\over J_r+|L_z|+J_z}\right)\cr
B&=4.5-3.5\tanh^2\left({|L_z|+J_z\over J_r+|L_z|+J_z}\right)
\end{split}\]
 The
quantities $\kappa$ and $\Omega_\phi$ above are epicycle and azimuthal
frequencies in the self-consistent, isolated spherical model evaluated at the
radius $R_\mathrm{c}$ of a circular orbit with angular momentum
\[\label{eq:defJtot}
J_{\rm tot}\equiv J_r + |J_\phi| + J_z.
\]
We take the argument of $\Rc$ to be $J_{\rm tot}$ to make it an approximate
function of energy, so $R_\mathrm{c}$ does not become small, and the epicycle
frequencies large, for stars on eccentric and/or highly inclined orbits with
small $|J_\phi|$.

The \df\ vanishes for $h>h_{\rm max}=10^6\kms\kpc$, and for $h<h_{\rm max}$
\begin{equation}
 f(\vJ) = \frac{N}{J_0^3}\,\frac{[1 + J_0 / ( J_\mathrm{core} +
 h(\vJ))]^{5/3}}{[1 + h(\vJ)/J_0]^{2.9}}-f_{\rm tide},
\end{equation}
where $N$, $J_0$, $J_{\rm core}$ and $f_{\rm tide}$ are constants. The
normalising constant $N$ determines the virial mass of the dark halo, $J_0$ sets the
radius of the transition between the inner and outer power-law segments of
the system's radial density profile, $J_{\rm core}=1.25\times10^{-2}\kms\kpc$
is a small number required to keep the central phase-space density finite and
$f_{\rm tide}$ is chosen to make $f$ vanish for $h= h_{\rm
max}=10^6\kms\kpc$.  Appendix C of PPB explains the rationale for this choice
of the dark halo's \df.

Once we have obtained a self-consistent model of an isolated halo, we freeze
the dependence of $\Rc$ on its argument, so while we relax $\Phi_{\rm tot}$
onto the potential that is jointly generated by all the Galaxy's components,
the dark halo's \df\ stays exactly the same function of $\vJ$. It is
essential to freeze the function $f(\vJ)$ during the introduction of the
discs and the bulge if one seeks to learn how the halo is distorted by the
gravitational fields of its companions.
\subsection{The bulge/bar and gas disc}

Our modelling technique restricts us to axisymmetric models, so we cannot use
a sophisticated model of the bulge/bar. Moreover, the data we use are only
sensitive to the bulge's contribution to radial forces. Therefore we do not
represent the bulge by a \df\ $f(\vJ)$ but by a fixed axisymmetric mass
distribution.  Following \citet{McMillan2011} we use a model similar to that
constructed by \citet{Bissantz2002}.

\begin{table}
 \centering
 \caption{Parameters for gas disc and bulge. With the exception of $\Sigma_0$
 and $R_\d$, these parameters were fixed.}
 \label{tab:mass_model_params}
 \begin{tabular}{lcl}
  \hline\hline
  Parameter & value & unit \\
  \hline
  Gas disc\\
  $\Sigma_0$ & 	$57.8$&$\msun\pc^{-2}$\\
  $R_\mathrm{d}$ &	$2R_\d({\rm stars})$&\\
  $z_\mathrm{d}$ &  $0.04$&$\kpc$\\
  $R_\mathrm{hole}$ & $4$&$\kpc$\\
  $M(\infty)$     & $17.7\times10^9$&$\msun$\\
  $M(R_0)$        & $3.53\times10^9$&$\msun$\\
  \hline
  Bulge\\
  $\rho_\mathrm{0,b}$	& $94.9$&$\msun\pc^{-3}$\\
  $r_\mathrm{0,b}$ & $0.075$&$\kpc$\\
  $r_\mathrm{cut,b}$& $2.1$&$\kpc$\\
  $q_\mathrm{b}$	&	 0.5& \\
  $\gamma_\mathrm{b}$	&	 0&\\
  $\beta_\mathrm{b}$	&	 1.8&\\
  $M(\infty)$      & $8.56\times10^9$&$\msun$\\
  \hline
 \end{tabular}
\end{table}

The density distributions of  the bulge is
\begin{equation} \label{eq:rho_spheres}
 \rho(R,z) = \frac{\rho_0}{m^\gamma(1+m)^{\beta-\gamma}}
 \exp[-(mr_0/r_\mathrm{cut})^2],
\end{equation}
where
\begin{equation}
 m(R,z) = \sqrt{(R/r_0)^2 + (z/qr_0)^2}.
\end{equation}
 Our model bulge has an axis ratio $q=0.5$ and extends
to $r_{\rm cut}=2.1\kpc$: Table~\ref{tab:mass_model_params} lists all the
parameters.

The gas disc is likewise represented by an axisymmetric distribution of
matter that has density
\begin{equation} \label{eq:rho_disc}
 \rho(R,z) = \frac{\Sigma_0}{2z_\mathrm{d}}
    \exp\left[-\left(\frac{R}{R_\mathrm{d}} + \frac{|z|}{z_\mathrm{d}} +
    \frac{R_\mathrm{hole}}{R}\right)\right].
\end{equation}
A non-zero parameter $R_\mathrm{hole}$ creates a central cavity in the disc.
The values of the parameters are given in Table~\ref{tab:mass_model_params}.
The scale length $R_\d$ is set equal to twice the scale length of the stellar
disc, and the surface density normalisation is adjusted to maintain the ratio
$13.5:35.5$ between the gas and stellar surface densities at $R_0$ that is
given in \cite{Flynn2006}.  $R_\d$ and $\Sigma_0$ are the only parameters
that are varied: the other parameters are fixed at the values adopted by P14
and earlier investigators.
\section{Fitting algorithm}\label{sec:algorithm}
Since the scale action $J_0$ of the dark halo's \df\ has little bearing on
the contribution of the dark halo to forces on stars in the inner Galaxy
($r\le10\kpc$), we fix $J_0$ to a value, $J_0=6000\kms\kpc$, consistent with
values obtained in preliminary tests of a procedure that did involve varying
$J_0$. Fixing $J_0$ effectively constrains the virial mass of the dark halo,
so there is then no need to impose an explicit upper limit on the halo mass
within $50\kpc$ as P14 did.  With $J_0$ determined, the only adjustable
parameter for the dark halo is its overall normalisation.

The disc \df\ has in principle 12 parameters: for each sub-disc a mass, the
normalising velocity dispersions $\sigma_{r0}$ and $\sigma_{z0}$, and the radial
scale lengths $R_\d$, $R_{\sigma_r}$, $R_{\sigma_z}$ of the mass density and
the velocity dispersions. We reduce the free parameter count to 9 by
assuming that $R_\d$ is the same for both discs, and for the thin disc setting
$R_{\sigma_r}=R_{\sigma_z}=2R_\d$. Since the radial range covered by RAVE
increases with $|z|$, the data do contain information about the thick disc's
values of $R_{\sigma_r}$ and $R_{\sigma_z}$, so we let these parameters be
chosen by the data.  Hence the disc \df\ requires 9 parameters (two masses,
four pseudo velocity dispersions, the radial mass scale length, two radial
scale lengths for the thick disc's velocity dispersions).

When fitting to the data we have in all 10 free parameters: 9 for the stellar
discs, and one for the dark halo.

The construction of a fully self-consistent model as described by PPB is
computationally quite expensive because for each trial potential the density must be
computed at every point of an extensive grid by integrating over all three
components of velocity. Consequently, it proved impractical to go through the
whole procedure in each iteration of the fitting process with in total 10
fitting
parameters. To speed up the process, we exploit the fact that we already know
in some detail the final mass distribution of the stellar component, because
(i) our disc \df\ always results in a density distribution very close to a
double-exponential and (ii) the star-count data from \citetJuric\ set
very tight constraints on the vertical structure of the disc. Hence for large
parts of the fitting process we do not deal with the disc \df\ directly, but
use a dummy potential of a double-exponential distribution with the right
vertical density structure. Hence our procedure is a follows:

\begin{enumerate}
\item[1.] We choose a normalisation of the total mass of the dark halo.

\item[2.] We adopt a double-exponential disc that has
the right vertical structure (satisfies the \citetJuric\ data).

\item[3.] We find the self-consistent equilibrium of our chosen dark halo in the
presence of the bulge, the gas disc and our adopted double-exponential
stellar disc.

 \item[4.] By adjusting masses and radial scale(s) of the double-exponential
disc (and consequently of the gas disc, which is tied to the stellar disc),
we seek satisfaction of the constraints on $\vc(R)$ listed in
Section~\ref{sec:constraints} -- these comprise the terminal velocities and data
for the masers and Sgr A*. Each time the the disc parameters are changed, we
relax the halo again in the presence of the updated disc. Thus after each
unsuccessful comparison with data of the rotation curve generated by the
dark-halo \df\ in the presence of the current double-exponential disc, we
return to step 2.

\item[5.] We choose a \df\ for the stellar disc that has the scale lengths
found in Step 4 and we adopt plausible values for the velocity-dispersion
parameters. 

\item[6.] Then we determine a new overall potential as follows. (i) We update
the contribution of the disc to the density that generates the potential to
the density obtained by integrating the disc \df\ over velocities using the
current estimate of the potential. (ii) We solve for the potential generated
by the updated disc density and the current estimate of the dark-halo
density. (iii) We update the contribution of the dark halo to the density
that generates the potential to the density obtained by integrating the halo
\df\ over velocities in the current potential. (iv) We again solve for the
overall potential.  Then we return to step (i) until the updates become
insignificant.

\item[7.] With the potential frozen at its current value, we adjust the
velocity-dispersion parameters in the disc \df\ to obtain a good fit of the
model's kinematics to the kinematics of the RAVE giants in eight spatial
bins. At this stage we include the contribution of the stellar halo with its
\df\ normalised as described in Section~\ref{sec:starhalo}. Then we return to
Step 6 and follow it with Step 7 until updates become negligible.

\item[8.] We compute the residuals between the vertical stellar density
profile of \citetJuric\ and that implied by the \df s of the disc and stellar halo.

\item[9.] If these residuals are unsatisfactory, we choose a new mass for the
dark halo and return to Step 2.

 \end{enumerate} 

This algorithm derives from two physical principles.  First the predicted
rotation curve is sensitive to the scale lengths of the disc components and
insensitive to the velocity dispersion parameters. Second the
velocity-dispersion parameters control the kinematics of stars within each
spatial bin with little sensitivity to the potential used. Consequently, even
when the potential employed is far from the truth, the RAVE data yield values
for the velocity-dispersion parameters that are close to the final best-fit
values. Given the value of $\sigma_{z0}$, the correct potential can be
identified by comparing the density profile predicted by the \df\ to the
star-count data.

Replacing the double-exponential discs by the density distribution generated
by the disc \df\ in step 6 changes the potential very little. We do not include the
stellar halo in the sources of the gravitational potential because its mass
should be negligible, even though the mass implied by our \df\ can be
significant depending on its ill-constrained behaviour at small actions that
do not contribute to the RAVE data.  Because the density distribution
generated by the disc \df\ is not \emph{exactly} the same as that generating
the dummy potential, it is not practical to set the \df\ normalisation such
that we obtain the same total mass as generates the dummy potential. Instead
we normalise such that the disc \df\ yields the same radial force at the Sun
as does the dummy potential.
\section{Results}\label{sec:results}
Here we describe our best-fit model. Table~\ref{tab:haloDF} gives the values of
the parameters in the final dark halo's  \df, and below the line some
derived properties. Its local density is
\begin{equation}
 \rhodm(R_0,0) = 0.01307 \msun\pc^{-3}=0.50\,\hbox{GeV}\cm^{-3}.
\end{equation}

\begin{figure}
\centering
\includegraphics[width=\hsize]{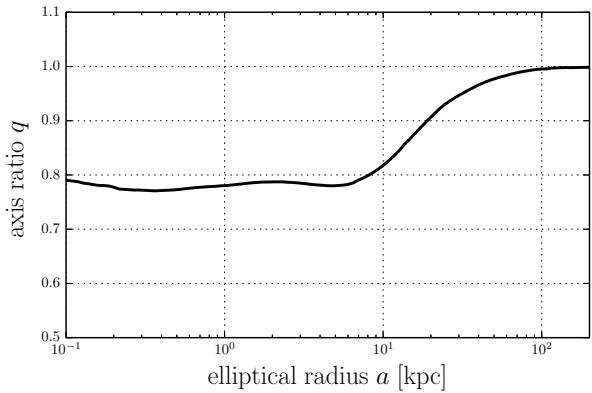}
\caption{The axis ratio $q$ of the final dark halo as a function of
semi-major axis length $a$.}
\label{fig:halo_q}
\end{figure}

In the absence of the discs and bulge, the \df\ of the dark halo generates a
spherical mass distribution. Fig.~\ref{fig:halo_q} shows the final dark
halo's axis ratio $q$ as a function of semi-major axis length $a$.  At
$a\la R_0$ the axis ratio is fairly constant at just below $0.8$, and then,
as the influence of the disc wanes, it increases towards $q=1$, implying
spherical symmetry. The axis ratio $q\simeq0.8$ in the inner Galaxy is consistent
with the findings of PPB, but their halo became rounder sooner because their
disc was more compact.

For the halo scale action $J_0$ we find a value around $6000\kms\kpc$.  This
parameter is, however, only weakly constrained by our data and strongly
correlated with the derived halo virial mass because the latter is largely
determined by the density of the dark halo well outside $R_0$, while our data
constrain the halo density at $r\lta R_0$. In fact, the virial mass of the
dark halo can be increased without much impact on our data by increasing
$J_0$ so the density within $R_0$ changes little.  Taken in isolation, i.e.
used to recover the dark halo's structure before the baryons fell in, our
dark-halo \df\ yields an NFW profile that has scale length $r_s\simeq16\kpc$,
virial radius $R_{200} = 223 \kpc$, and virial mass $M_{200} = 1.4 \times
10^{12}\msun$.

The stellar discs have a scale length $R_\d=3.66 \kpc$ and total mass $3.6
\times 10^{10} \msun$. At $R_0$ the stellar surface density is
$46.3\msun\pc^{-2}$. The total baryonic mass (i.e., including the Bulge and
the gas disc) is $6.2 \times 10^{10} \msun$. Hence we have a (visible) baryon
fraction of 4.2 per cent. The remaining parameters of the disc \df\ are given
in Table~\ref{tab:discDF}.

\begin{table}
 \centering
 \caption{Parameters of the stellar disc \df. All parameters were adjusted
 when fitting the data.}
 \label{tab:discDF}
 \begin{tabular}{rcl}
  \hline\hline
  Parameter & value & unit \\
  \hline
  $\Sigma_0$ & $110$ & $\msun\pc^{-2}$ \\
  $R_{\rm D}$ & 3.66 & $\kpc$ \\
  Thin disc \\
  $\sigma_{R,\rm thn}$ & 35.40 & $\kms$ \\
  $\sigma_{z,\rm thn}$ & 26.00 & $\kms$ \\
  $R_{\sigma_R,\rm thn}$ & $2R_\d$ & \\
  $R_{\sigma_z,\rm thn}$ & $2R_\d$ & \\
  \hline
  Thick disc \\
  $\sigma_{R,\rm thk}$ & 52.78 & $\kms$ \\
  $\sigma_{z,\rm thk}$ & 53.33 & $\kms$ \\
  $R_{\sigma_R,\rm thk}$ & 11.6 & $\kpc$ \\
  $R_{\sigma_z,\rm thk}$ & 5.01 & $\kpc$ \\
  $F_{\rm thk}$ & 0.416 \\
  \hline\hline
 \end{tabular}
\end{table}

\begin{table}
\centering
\caption{Parameters of the dark halo's \df. Of these parameters, only $N$ was
adjusted when fitting the data. $R_{200}$ and $M_{200}$ are derived
parameters of the dark halo that is generated by the \df\ in the absence of other
components.}
 \label{tab:haloDF}
\begin{tabular}{rcl}
\hline\hline
Parameter & value & unit\\
\hline
$N$ & $8.825\times10^{12}$ & $\!\msun$\\
$h_{\rm max}$ & $10^6$ & $\!\kms\kpc$\\
$J_0$ & 6000 &$\!\kms\kpc$\\
$J_{\rm core}$ & $1.25\times10^{-2}$ &$\!\kms\kpc$\\
\hline
$R_{200}$ & 223 & $\!\kpc$\\
$M_{200}$ & $1.4\times10^{12}$&$\!\msun$\\
\hline\hline
\end{tabular}
\end{table}
\begin{figure}
 \centering
 \includegraphics[width=0.47\textwidth]{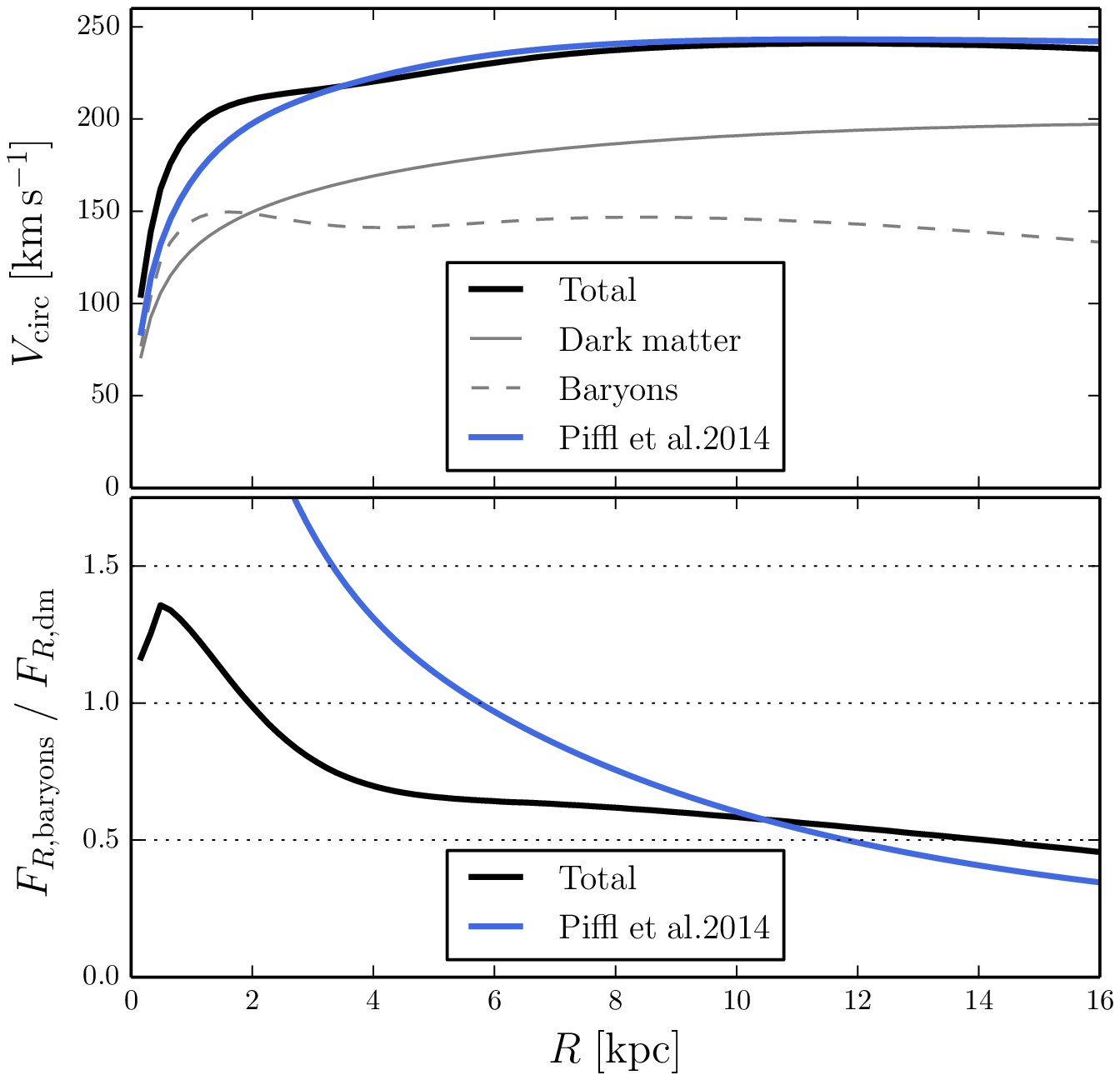}
 \caption{Upper panel: $\vc(R)$ for our final Galaxy model (black lines)
 and, for comparison, for the P14 model (blue lines). The solid lines show
 the total $\vc$, while the dashed and dotted lines show the contributions
of the dark halo and the baryons, respectively. Lower panel: the ratio of the
radial forces stemming from the baryonic components and the dark matter halo
for our model and the P14 model.}
 \label{fig:rotationcurve}
\end{figure}
In the upper panel \figref{fig:rotationcurve} we show the final model's circular
speed $\vc(R)$.
For comparison we added in blue $\vc(R)$ for the model of P14
with $q = 0.8$. The curves are very similar except for the region $R\la3\kpc$
in which circular orbits are impossible on account of the Galactic bar.
Consequently, the model is unconstrained in this region. The grey curves in
this panel show circular speeds that are generated by the dark matter (full
curve) and baryons (dashed curve). The lower panel of
\figref{fig:rotationcurve} shows the ratio of the radial forces coming from
the baryons and dark matter. On account of adiabatic contraction, our model
contains much more dark matter at small radii than that of P14. At the Sun
the ratio, $0.61$, is even lower than that, $0.85$, of P14, so only 38 per cent
of the radial force on the Sun is due to baryons. Moreover, whereas in the
P14 model the contribution to $F_R$ from baryons rises steeply inwards,
becoming dominant at $R\lta5\kpc$, in the present model it remains low until
$R\simeq3\kpc$ and the baryons are dominant only at $R\lta2\kpc$.

\begin{figure*}
 \centering
 \includegraphics[width=0.7\textwidth]{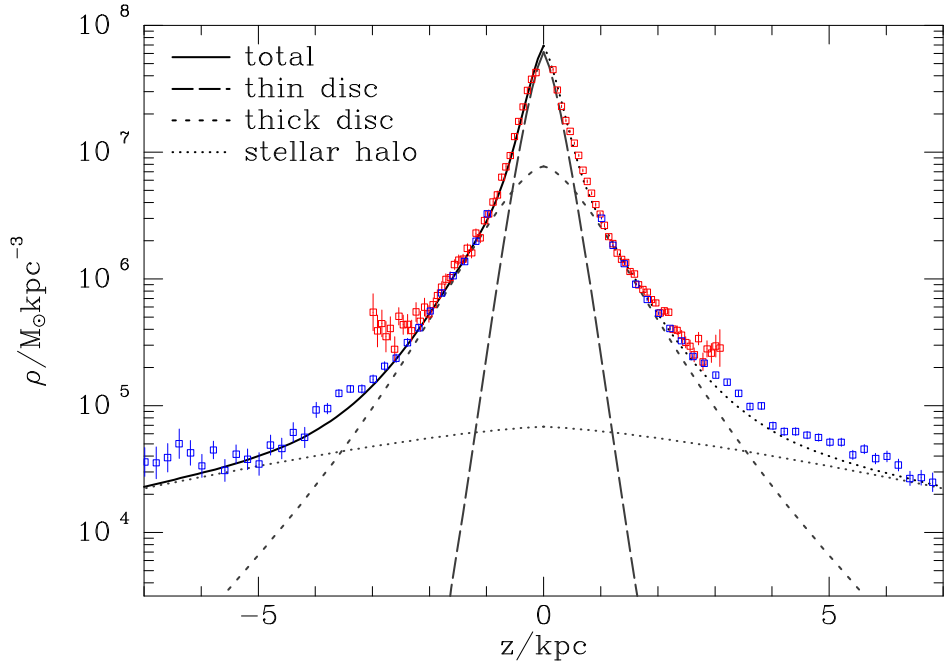}
 \caption{Vertical stellar density profiles in the solar annulus of our
composite model (solid line), the thin disc (dashed line) and the thick disc (dotted
line), and the stellar halo (dotted line). The red and blue error
bars show the data from \citet{Juric2008}. Only the data represented by the
red error bar was used in the fitting process.}
 \label{fig:verticalprofile}
\end{figure*}

\figref{fig:verticalprofile} compares the vertical stellar density profile
above the Sun with the star count data from \citetJuric. The model star
density comprises that predicted by the disc \df\ plus that predicted by the
\df\ of the stellar halo when normalised as described in
Section~\ref{sec:starhalo}.  The agreement between model and data is
excellent. 

\begin{figure*}
 \centering
 \includegraphics[width=0.98\textwidth]{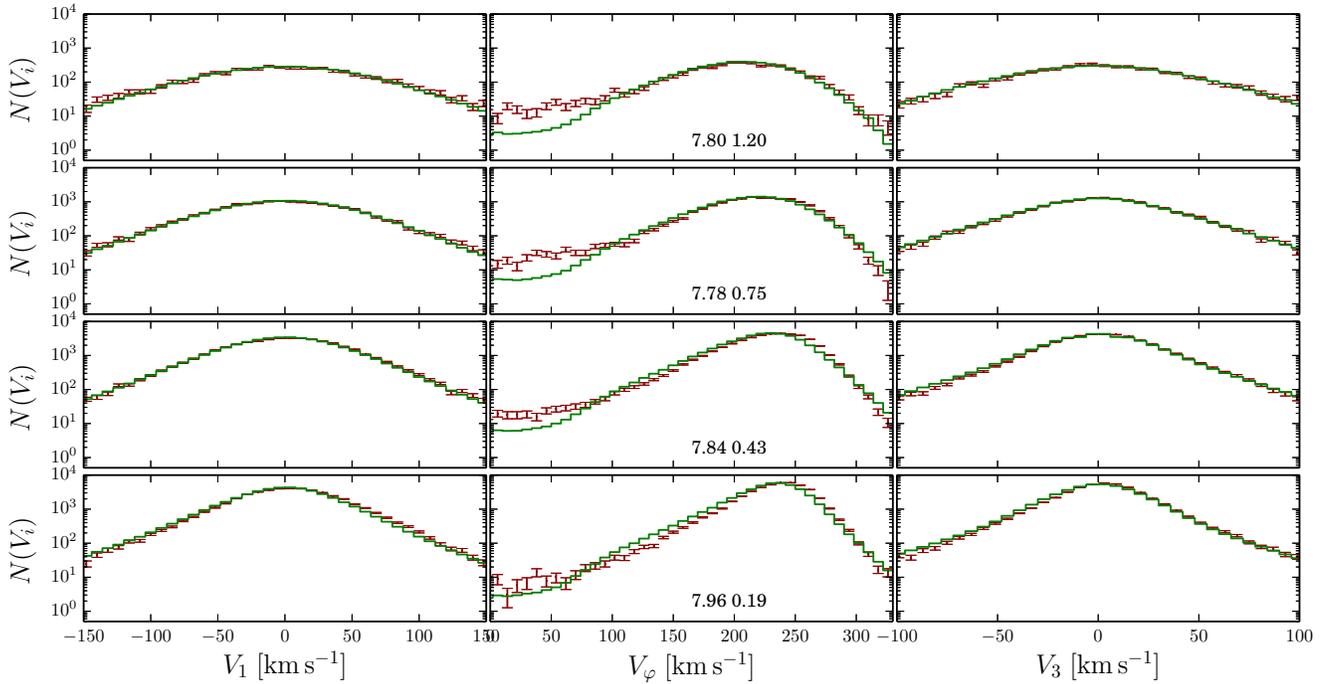}
 \caption{Model (green curves) and observed (red error bars) velocity
distributions for RAVE giants with $R_0-1\kpc<R<R_0$. Each row represents a
slice in distance from the Galactic plane. The $(R,z)$ coordinates of the
barycentre of the stars in each slice are given in the middle panels.  These
are also the locations where the \df\ was evaluated.  A correction described
in PPB was applied to the model prediction to incorporate the influence of
errors in the measurement of both velocities and distances.}
 \label{fig:RAVE}
\end{figure*}

\figref{fig:RAVE} shows our fit to the kinematics of RAVE giants in the four
spatial bins that lie inside $R_0$: the coordinates of the barycentre of each
bin are given at the bottom of the centre panel of each row. The fits are
excellent apart from a deficit of almost non-rotating stars that becomes more
marked as one moves away from the plane. This deficit points to weakness in
our choice of \df\ for the stellar halo and/or the thick disc.

\begin{figure}
 \centering
 \includegraphics[width=0.47\textwidth]{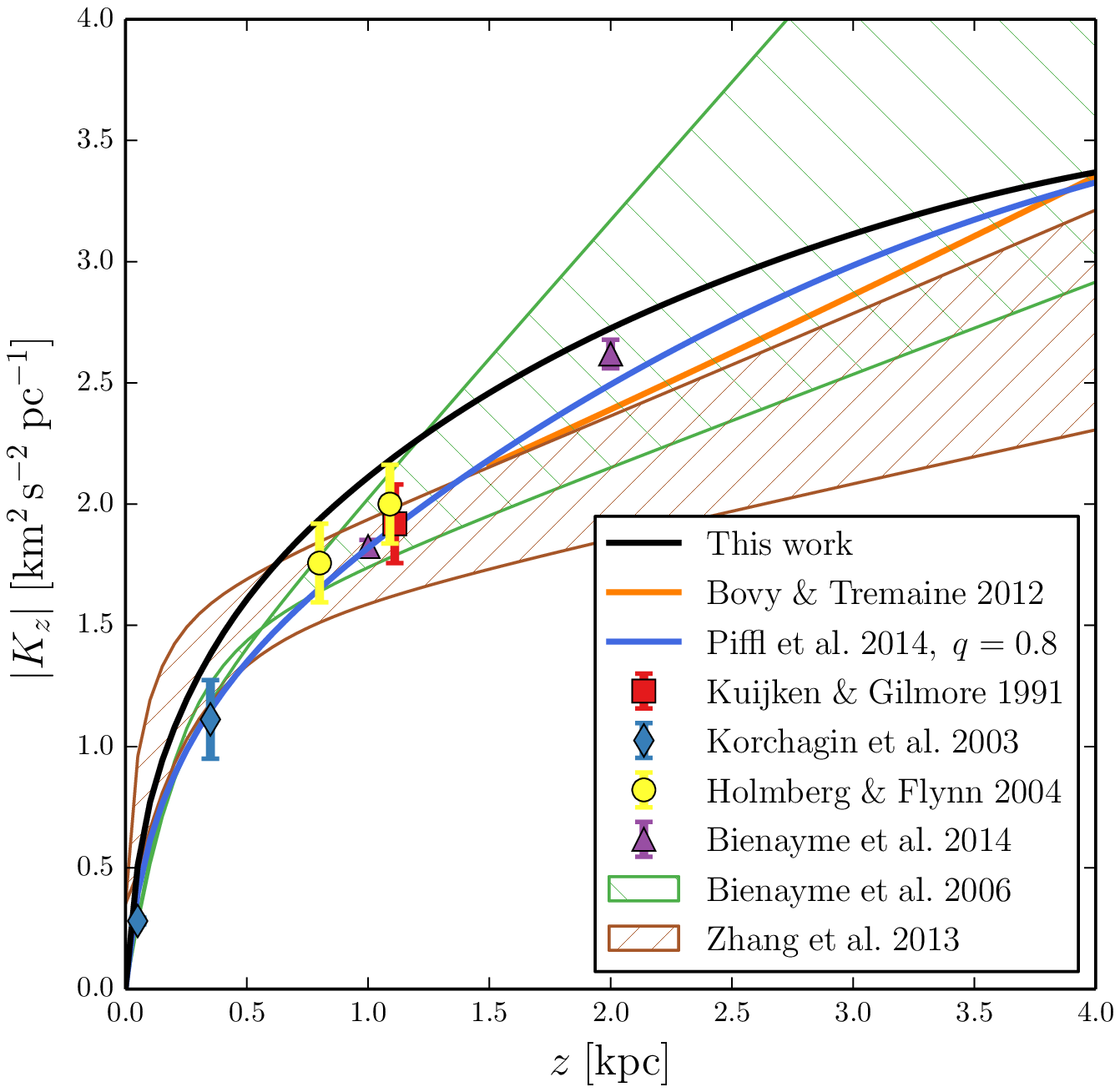}
 \caption{Vertical force $|K_z|$ in the solar cylinder as a function of
distance $z$ from the Galactic plane. See the main text for a further
discussion}
 \label{fig:Kz}
\end{figure}
The black curve in \figref{fig:Kz} shows the vertical gravitational force
$K_z$ in the solar cylinder as a function of distance from the Galactic
plane, $z$. The brown and blue curves show force laws computed from
\cite{BovyTremaine2012} and P14, respectively, the blue curve being that for
the P14's model with $q=0.8$. Error bars show results from a number of other
studies, including the seminal work of \cite{Kuijken1991}. Only
\cite{Korchagin2003}, \cite{Bienayme2006} and P14 constrain $K_z$
significantly at $|z|<1\kpc$.  In general there is good agreement between the
studies. 

Even though our disc is slightly less massive than that of
P14 ($3.6$ rather than $3.7\times10^{10}\msun$), it generates larger $K_z$
at all $z$ because its longer scale length implies a higher surface density
at the Sun ($46.3$ rather than $37.1\msun\pc^{-2}$). In fact, its high local
surface density ensures that at all values of $z$ its $K_z$ exceeds that from
other studies.
\begin{figure}
\centerline{\includegraphics[width=.45\textwidth]{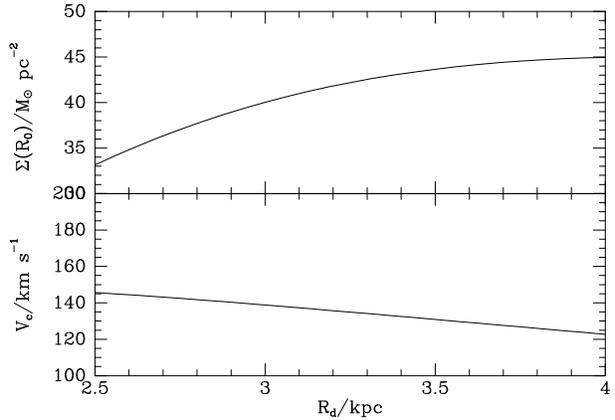}}
\caption{Lower panel: contribution to $\vc(R_0)$ from a thin exponential disc
with mass $M_{\rm disc}=3.6\times10^{10}\msun$ and varying scale lengths.
Upper panel, the surface density of such a disc at $R_0$.}\label{fig:exptl}
\end{figure}

\section{Discussion}\label{sec:discuss}

By replacing the fixed NFW dark halo used by P14 with a dynamically active
dark halo, one increases the ratio of dark-halo densities, $\rhodm(3\kpc)/\rhodm(R_0)$.
Since the value of $\rhodm(R_0)$ is strongly constrained by solar
neighbourhood data -- which leave no doubt that the disc can contribute only
part of  $\vc(R_0)$ -- the tendency encountered by PPB for a more centrally
concentrated dark halo to drive $\vc(3\kpc)$ above the observational limits
must be addressed by increasing the scale length of the disc. If done at
fixed disc mass, this operation moves disc material outwards, thus
counteracting the increased central concentration of the dark halo on its
becoming active.  However, as the lower panel of Fig.~\ref{fig:exptl} shows,
increasing $R_\d$ at fixed disc mass reduces the disc's contribution to
$\vc(R_0)$. Hence there is little scope for reducing either $M_{\rm disc}$ or
$\rhodm(R_0)$ below the values found by P14. The upper panel of
Fig.~\ref{fig:exptl} shows that as $R_\d$ is increased at fixed $M_{\rm
disc}$, the local stellar surface density $\Sigma(R_0)$ increases. The scope
for such an increase is limited by the RAVE and \citetJuric\ data.
Hence the present optimisation process leads to a model with a longer disc scale
length but similar mass and $\rhodm(R_0)$. Pressure
from the constraints on $\vc(3\kpc)$ drives the disc to higher values of
$\Sigma(R_0)$ and $K_z$ than are ideal from the perspective of the RAVE and
\citetJuric\ data.

\subsection{Constraints from microlensing}

Above we found a model with an adiabatically compressed dark halo that is
consistent with all the observational constraints of
Section~\ref{sec:constraints}.  However, we now show that this model is not
consistent with data not so far considered, namely the measured optical depth
to microlensing of bulge stars.  \citet{BinneyEvans2001} took the local
density of dark matter to be $\rhodm(R_0)=(0.0136\pm0.007)\msun\pc^{-3}$
in close agreement with our value. For any adopted value of the local
dark-matter density, the local surface density of the baryonic disc followed
from the measured total surface density $\Sigma_0=71\pm6\msun\pc^{-2}$ within
$1.1\kpc$ of the plane \citep{Kuijken1991}. Binney \& Evans assumed that the
baryons were arranged in such a way, as regards the ellipticity of the bar
and the scale height of the disc, that the optical depth to microlensing
along the line of sight $(l=0,|b|\simeq4^\circ)$ was maximised for any given
contribution to $\vc$.  Taking the optical depth on this line of sight to be
at least $2\times10^{-6}$ \citep{Popowski2005}, and extrapolating the adopted
local dark-matter density inwards as a power law of slope $-\alpha$, they
showed that the inferred values of $\vc$ exceeded those implied by
measurements of tangent velocities unless $\alpha$ was significantly smaller
than unity or $\rhodm(R_0)<0.007\msun\pc^{-3}$. Our adiabatically
compressed dark halo has $\rhodm(R_0)=0.014$ and $\alpha>1$, so it comfortably
violates these constraints.

\begin{figure}
\includegraphics[width=\hsize]{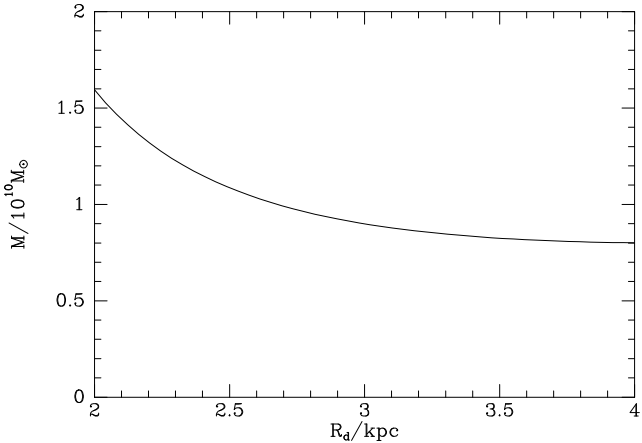}
\caption{Mass as a function of scale
length for a thin exponential disc that contributes a given amount,
$10\msun\pc^{-2}$, to the local surface density.}\label{fig:MoS}
\end{figure}

\subsection{Are long disc scale lengths viable?}

To simplify the fitting process, we set the scale lengths of the thin and
thick discs equal, and obtained $R_\d=3.66\kpc$. This value is in good
agreement with that obtained for the thin disc from SEGUE data by
\cite{Bovy2012c}. It is also consistent with the long scale length
($R_\d=3.6\kpc$) \citetJuric\ measured for the thick disc.

However, the thick disc is now generally distinguished from the thin disc by
high values of the abundance of $\alpha$ elements such as Mg relative to Fe
at a given value of [Fe/H]. When the thick disc is defined thus, current data
suggest that the thick disc has a shorter scale length than the thin disc
\citep{Hayden2015,RecioBlanco2014,Nissen2014}, contrary to the finding of
\citetJuric.  If the thick disc does have a short scale length, this will
have a significant impact on the viability of the current model, because the
model assigns significant mass to the thick disc ($1.06\times10^{10}\msun$).
The vertical density profile of \citetJuric\ essentially fixes the
contribution of the thick disc to the local surface density.
Fig.~\ref{fig:MoS} shows as a function of scale length the mass of an
exponential disc that contributes $10\msun\pc^{-2}$ to the local surface
density. It shows 
that shortening the scale length of the thick disc will make it more
massive. Its peak contribution to $\vc$ will be increased by both this
increase in its mass and the shortening of its scale length. Moreover, any
increase in the central density of the thick disc will further compress the
dark halo and further increase $\vc$.  Hence, it seems unlikely that there is
significant scope for reducing the scale length of the thick disc without
violating the constraints of Section~\ref{sec:constraints}.

There are indications that the chemically defined, low-$\alpha$ abundance,
thin disc is flared \citep{Minchev2015}. This being so, our geometrically thick disc
with a long scale length may be consistent with a short scale length for the
$\alpha$-enhanced thick disc. Consequently, the requirement for a large scale length
is not as serious an issue as the requirement for a higher optical depth to
microlensing.

\subsection{Selection functions}

An aspect of the present work which should be improved, is the neglect of
selection functions. Any survey captures only a fraction of the stars that
exist within any volume. The fraction that is captured varies with distance,
and with intrinsic stellar properties such as mass, chemistry and age, that
determine the star's absolute magnitudes in the relevant wavebands.  The
stellar densities from \citetJuric\ on which we have relied take fully into
account the selection function of the SDSS photometric survey. In comparing
the kinematics predicted by models with the RAVE data we should properly have
considered biases towards younger or older stars, since the age of a star
affects its likely kinematics. \cite{SandersBinney2015a} explored the impact
of accounting for age biases on the best-fitting parameters of \df s of the
present type in the context of GCS data. This exercise has not yet been
performed for RAVE data, but it should be. We doubt that proper treatment of
age biases would materially affect the conclusions we have reached here.

\section{Conclusions}\label{sec:conclude}
We have extended the approach of PPB from the construction of a single Galaxy
model in which both the stellar disc and the dark halo are represented by
distribution functions in the the self-consistently generated gravitational
potential to a systematic search through a multi-dimensional parameter space
of such models. As in the model of PPB the dark halo
has the structure expected if the baryons fell in smoothly, so the dark
matter was adiabatically compressed. Consequently, the dark halo is more
centrally concentrated than the NFW profile, which arises in simulations of
cosmological clustering of dark matter only.

Whereas the model of PPB violated the observational constraints on $\vc(R)$
at $R<R_0$, our search of parameter space has identified a model that is
consistent with these constraints. The secret to achieving consistency is an
increase in the scale length of the disc at essentially fixed stellar mass.
The derived disc scale length, $R_\d=3.66\kpc$, may be acceptable for the
thin disc, but it is probably unacceptably large for the thick disc.
However, our work suggests that a model with an adiabatically compressed dark
halo in which the thick disc has a realistically short scale length cannot be
made consistent with the constraints on $\vc$ at small radii.

Since dark matter does not cause microlensing and the model assigns to the
dark halo much of the overall density at small radii, it violates by some
margin the constraints on the dark halo that \cite{BinneyEvans2001} deduced
from the microlensing data.

Hence by ruling out adiabatic compression this study furnishes compelling
evidence that there has been a significant transfer of energy from the
baryons to the dark halo.  Such a transfer has been proposed by many authors
because it can arise through drag on the bar
\citep{TremaineWeinberg1984,Sellwood2008}, drag on molecular clouds
\citep{NipotiBinney2015}, energy injection by supernovae
\citep{Mashchenko2008}, and by cosmic infall \citep{Naab2009,Goerdt2010}.

Given that we can now exclude adiabatic compression of the dark halo, the
natural next step is to impose priors on the scale lengths of the thin and
thick discs derived from spectroscopic studies of Galactic chemistry, and to
add the microlensing data to the constraints listed in
Section~\ref{sec:constraints}.  Then one should use the fitting procedure
developed here to build self-consistent models using \df s for the dark halo
that imply increasing extents of central heating of dark matter by baryons.
In this way it should be possible to place a lower limit on extent of
dark-matter heating. We hope to publish the results of such a study in the
near future.

\section*{Acknowledgements}
The research leading to these results has received funding from the European Research
Council under the European Union's Seventh Framework Programme (FP7/2007-2013)/ERC
grant agreement no.\ 321067.

Funding for RAVE has been provided by: the Australian Astronomical Observatory;
the Leibniz-Institut f\"ur Astrophysik Potsdam (AIP); the Australian
National University; the Australian Research Council; the French National Research
Agency; the German Research Foundation (SPP 1177 and SFB 881); the
European Research Council (ERC-StG 240271 Galactica); the Istituto Nazionale
di Astrofisica at Padova; The Johns Hopkins University; the National Science
Foundation of the USA (AST-0908326); the W. M. Keck foundation; the Macquarie
University; the Netherlands Research School for Astronomy; the Natural
Sciences and Engineering Research Council of Canada; the Slovenian Research
Agency; the Swiss National Science Foundation; the Science \& Technology
Facilities Council of the UK; Opticon; Strasbourg Observatory; and the
Universities of Groningen, Heidelberg and Sydney. The RAVE web site is at
http://www.rave-survey.org.

\bibliographystyle{mn2e}
\bibliography{all_references}
\label{lastpage}
\end{document}